\documentclass[10pt,a4paper]{article}
\usepackage{amsmath}
\usepackage{amsfonts}
\usepackage{amssymb}
\usepackage{graphicx}
\input{epsf}
\newcommand{\be}{\begin{eqnarray}}
\newcommand{\ee}{\end{eqnarray}}
\usepackage{epsfig}
\usepackage{cite}
\usepackage{textcomp}
\title{Self similarity of the dark matter dominated objects
and the shape of small scale power spectrum}
\author{M. Demia\'nski$^{a,b}$,\footnote{Marek.Demianski@fuw.edu.pl}\,
A. Doroshkevich$^{c,d}$\footnote{dorr@asc.rssi.ru}\\[15pt]
$^{a}$Institute of Theoretical Physics, University of Warsaw,
02-003, Warsaw Poland\\
$^{b}$Department of Astronomy, Williams College, Williamstown,
MA 01267, USA\\
$^{c}$Lebedev Physical Institute of Russian Academy of Sciences,
117997, Moscow, Russia\\
$^{d}$National Research Center Kurchatov Institute, 123182,
Moscow, Russia}

\begin{document}
\maketitle
\begin{abstract}
We analyzed available observational data of a sample of dark matter (DM) dominated galaxies and
clusters of galaxies  and we have found correlations between
the virial mass, $M_{vir}$, of halos and basic parameters of
their cores, namely, the mean DM density, pressure and entropy.
These correlations are a natural consequence of similar evolution
of all such objects. It is driven mainly by gravitational
interactions what implies a high degree of self similarity
of both the process of halos formation and their internal
structure.

We confirmed the CDM--like shape of both the small and
large scale power spectrum. However, our reconstruction
of the evolutionary history of observed objects requires
either a multicomponent composition of DM or a more complex
primordial power spectrum of density perturbations with
significant excess of power at scales of clusters of
galaxies and larger. We demonstrated that a model with
suitable combination of the heavy DM particles (CDM) and
DM particles with large damping scale (HDM) could provide
a successful description of the observational data in a
wide range of masses.
\end{abstract}
{\it Keywords:} Cosmology: formation of DM halos, galaxies and
clusters of galaxies -- initial power spectrum -- composition
of dark matter

\section{Introduction}

One of the important goals of cosmology is to establish correlations
between the observed Universe --  the CMB, Large Scale Structure
(LSS), galaxies etc. -- and the processes that occurred at the
earlier epochs of evolution of the Universe and are encoded in
the initial power spectrum of density perturbations. For larger
scale $D> 10 Mpc$ this problem is approximately solved by the
CMB observations of the WMAP (Komatsu et al. 2011; Hinshaw et
al. 2013) and Planck (Ade et al. 2016) missions and is realized
as the standard $\Lambda$CDM model. However the shape of the
power spectrum at small scale remains unknown and its
investigation is one of the actual current problem of
cosmology. Now it is unfortunately not possible to  obtain
reliable information about this very important issue.
Next very interesting problem is the universality
and the self similarity of the internal structure of DM
dominated halos in a wide range of their masses and sizes.

To discuss these problems we use the improved quantitative
descriptions of observed characteristics of DM dominated
objects -- dSph and Ultra Diffuse (UDG) galaxies, and
clusters of galaxies -- for which we can expect the
relatively weak contribution of baryonic component. As
usual we will characterize such virialized objects by
their virial mass, $M_{vir}$, which is the most popular
basic characteristic of objects in spite of the low
precision of its observational determination. We also use
the well known correlation between the virial masses of
DM dominated objects and the period (or redshift) of their
formation. The DM dominated objects attract our attention as
their evolution is driven by gravitational forces only and
thus it is the simplest one as compared with the complex
evolution of numerous baryon dominated galaxies.

The relict concentration of baryons is moderate ($\sim
16\%$) and its influence on the object evolution is also
weak. However this influence is strongly enhanced for
the most abundant objects with virial masses $\sim 10^9\leq
M_{vir}/M_\odot\leq 10^{13}$ and virial velocities $\sim 15
km/s\leq v_{vir}\leq 400km/s$. For such objects the heating
of baryons by shock waves, subsequent cooling and infall
into central region strongly distort their internal
structure and lead to formation of  Irr, spiral and
elliptical galaxies. These processes are not so efficient
for dwarf galaxies with the low virial temperature
$T_{vir}\leq 1.5 \cdot 10^4 K$ and clusters of galaxies
with the high virial temperature $T_{vir}\geq 10^7 K$
and low density of baryons.

Evidently the halo formation is a deterministic process
and properties of virialized objects correlate with
properties of the initial perturbations. However, the
complex character of the process of halo formation (see,
e.g. Demia\'nski et al. 2011) destroys many correlations.
Simulations show that for the DM dominated objects
which evolve mostly due to gravitational
interactions the more stable characteristics are the
object rotation and the internal structure of their central
regions -- the mean density, pressure and entropy of their
cores. It can be expected that these characteristics
are moderately distorted in the course of halos formation
and evolution and they can be linked with properties of
initial perturbations.

Properties of both simulated and observed DM dominated
virialized objects are usually described in the framework
of spherical models such as the Navarro -- Frenk -- White
(NFW) proposal Navarro et al. (1997),
isothermal or Burkert (1995) models. This approach allows
to discuss and to link together both the general parameters
of a halo, such as its virial mass, period of its formation,
relations between the thermal and gravitational energy,
and its internal properties.
This approach allows to obtain a very simple, though
rough, general description of the process of DM halos
formation and introduces some hierarchy of formed objects.
To illustrate  problems of this approach we can refer to the papers of Weisz
et al. (2014) where reconstructions of evolution are presented
for several dwarf galaxies. High differences of these
evolutionary tracks emphasize the problems arising for
our analysis and explain the unavoidable scatter of
resulting estimates.

This analysis also  reveals a possible deviations of the
main observed characteristics from expectations based on
standard assumptions about the DM composition and/or
initial power spectrum used in theoretical models and
simulations. An example of such deviation is the 'To Big
to Fail' effect (Boylan--Kolchin et al. 2012; Garrison--
Kimmel et al., 2014a,b; Tollerud et al. 2014; Klypin et
al. 2015; Hellwing et al. 2015; Brook, \& Cintio 2015),
what indicates that the usually accepted models should
be improved.

Of course, our approach is the first and very rough
quantitative approximation. More extended discussion of
these problesms can be found in Demia\'nski\,\&\,
Doroshkevich (2015). Evidently these results will be
discussed and corrected many times as new observations
and special simulations will become available.

\subsection{Cosmological parameters}

In this paper we  consider the spatially flat $\Lambda$
dominated model of the Universe with the Hubble parameter,
$H(z)$, the mean critical density $\langle\rho_{cr}\rangle$,
the mean density of non relativistic matter (dark matter
and baryons), $\langle\rho_m(z)\rangle$, and the mean density
and mean number density of baryons, $\langle\rho_b(z)
\rangle\,\&\,\langle n_b(z)\rangle$, given by Komatsu et al.
2011, Hinshaw et al. 2013:
\[
H^{2}(z) = H_0^2[\Omega_m(1+z)^3+\Omega_\Lambda],\quad H_0=
100h\,{\rm km/s/Mpc}\,,
\]
\be
\langle\rho_b(z)\rangle = {3H_0^2\over 8\pi G}\Omega_b(1+z)^3
\approx 4\cdot 10^{-31}(1+z)^3\Theta_b\frac{g}{cm^3},\quad
\Theta_b=\frac{\Omega_bh^2}{0.02}\,,
\label{basic}
\ee
\[
\langle\rho_m(z)\rangle =
2.5\cdot 10^{-30}(1+z)^3\Theta_m\frac{g}{cm^3}=
34(1+z)^3\Theta_m\frac{M_\odot}{kpc^3},\quad \Theta_m=
\frac{\Omega_mh^2}{0.12}\,,
\]
Here $\Omega_m=0.24\,\&\,\Omega_\Lambda=0.76$ are the mean
dimensionless density of non relativistic matter and dark
energy, $\Omega_b\approx 0.04$ and $h=0.7$ are the
dimensionless mean density of baryons, and the dimensionless
Hubble constant measured at the present epoch. Cosmological
parameters presented in the recent paper of the Planck
collaboration (Ade et al. 2016) slightly differ from
those used above.

For this model with $\Omega_m\approx 0.25$ the evolution
of perturbations can be described with sufficient precision
by the expression
\be
\delta\rho/\rho\propto B(z),\quad B^{-1}(z)\approx
\frac{1+z}{1.35}[1+1.44/(1+z)^3]^{1/3}\,.
\label{bbz}
\ee
(Demia\'nski \& Doroshkevich, 1999, 2004, 2014; Demia\'nski
et al. 2011). For $z=0$ we have $B=1$ and for $z\geq 1,\,B(z)$
is reproducing the exact function with accuracy better than
90\%. For $z\geq 1$ these relations simplify. Thus, for
the Hubble constant and the function $B(z)$ we get
\be
H^{-1}(z)\approx \frac{0.85\cdot 10^{18}s}{\sqrt{\Theta_m}
(1+z)^{3/2}},\quad B(z)\approx \frac{1.35}{1+z}\,.
\label{bzz}
\ee

\section{Physical model of halos formation}

\subsection{DM density in virialized halos}

In this paper we assume that all relaxed DM halos
are described by the NFW density profile
\be
\rho(x)=\frac{\rho_0}{x(1+x)^2},\quad x=r/r_s\,,
\label{nfw-d}
\ee
where $\rho_0(M_{vir}),\,\&\,r_s(M_{vir})$ are model
parameters. Using this density we get that
\[
M(r)=M_sf_m(r/r_s),\quad M_s=4\pi\rho_0r_s^3,\quad
M_{vir}=M_sf_m(c)\,,
\]
\be
 f_m(x)=ln(1+x)-x/(1+x),\quad c=R_{vir}/r_s\geq 3\,,
\label{nfw-m}
\ee
\[
M(r_s)=M_sf_m(1)\approx 0.2M_s\,,
\]
where $c$ is the concentration and $R_{vir}$ is the halo
virial radius.  For the mean density of halos,
$\langle\rho_{vir}\rangle$ and the mean density of their
central core $\langle\rho_s \rangle$ we get
\[
\langle\rho_{vir}\rangle=3M(R_{vir})/4\pi R_{vir}^3=
3\rho_0f_m(c)/c^3\,,
\]
\[
\langle\rho_s\rangle=3M(r_s)f_m(1)/4\pi r_s^3\approx 0.6\rho_0\,,
\]
and finally we have
\be
\langle\rho_s\rangle=5.4\langle\rho_{vir}\rangle\left(\frac{c}{3}
\right)^3\frac{1}{f_m(c)}\,.
\label{nfw-ms}
\ee

These relations link together the fundamental characteristics
of halos, namely, their mean density, concentration and masses.
Moreover as the function $f_m(c)$ is a slowly varying function of $c$
than for the most interesting population of objects we can
neglect the differences between $M_{vir}$ and $M_s$:
\[
 4\leq c\leq 8,\quad 0.8\leq f_m(c)\leq 1.3\,,
\]
\be
M_{vir}\approx M_s(1\pm 0.25)\approx 5M(r_s)(1\pm 0.25)\,.
\label{mvir}
\ee
These relations allow to roughly estimate  the parameters
of observed objects in spite of scarcity of observational
data.

\subsection{The redshift of halo formation}

For each mass of DM halo its formation is a complex process
extended in time, what causes some ambiguity in the halo
parameters such as its virial mass and the epoch or redshift
of its formation (see, e.g. discussion in Partridge \& Peebles
1967a,b; Peebles 1980; and more recently Kravtsov \&
Borgani 2012). The main stages of this process can be
investigated in details with numerical simulations (see,
e.g. Demia\'nski et al. 2011). The analytic
description of this process is however problematic.

For a virialized DM halo of mass $M_{vir}$ the redshift (or
epoch) of its formation, $z_f$, and the corresponding mean DM
density $\langle\rho_{vir}\rangle$ were roughly estimated by
Partridge \& Peebles (1967a,b) and were later determined
more accurately in the framework of the simple {\it
phenomenological} toy models (Zel'dovich\,\&\,Novikov 1983;
Lacey \& Cole 1993; Bryan \& Norman 1998). According to
these models the virial density is proportional to the
mean density (\ref{basic}) at the moment of object formation,
\be 
\langle\rho_{vir}\rangle=18\pi^2\langle\rho_m(z_f)
\rangle\approx\rho_{200}(1+z_f)^3\,, 
\label{rmod}
\ee
\[
\rho_{200}=200\langle\rho_m(0)\rangle=
0.68\cdot 10^4M_\odot/kpc^3=5\cdot 10^{-28}g/cm^3.
\]
In the Lacey -- Cole model halos are considered as formed
at the moment of collapse of homogeneous spherical DM
clouds and they are described by the adiabatic Emden model
with $\gamma=5/3,\,n=3/2$ (Peebles 1980; Zel'dovich\,\&\,
Novikov 1983).

The advantage of this approach is its apparent universality
and reasonable results that are obtained for massive clusters
of galaxies. However, both its precision and range of
applicability are limited owing to noted simplified
assumptions. Thus, even the numerical coefficient $18\pi^2$
is determined by the Emden model for halo description.
An important, but usually ignored, special feature of the Lacey
- Cole model is the strong mass dependence of the virial density,
\be
\langle\rho_{vir}\rangle\propto M_{vir}^{-2},\quad
(1+z_f)\propto M_{vir}^{-2/3} \,.
\label{mvir}
\ee

Attempts to use these relations for unified description of
observed DM halos of galactic and clusters scales
immediately leads to unacceptable results. Thus,
expectations of the model (\ref{rmod},\,\ref{mvir}) are in
contrast with the weaker mass dependence of the observed
parameters of DM dominated objects (sections 4).
It is also important that for dSph galaxies Eq.
(\ref{rmod}) leads to a very high value of $z_f\sim 15$,
what exceeds both the age of dSph galaxies derived from
observations of stars (see, e.g., Weisz et al. 2014;
Karachentsev et al., 2015) and the redshift of reionization
$z_{reio}\sim 9$ determined by Planck (Ade 2016). Nonetheless
the main significance of the Lacey -- Cole model (\ref{rmod})
is the clear introduction of the concept of the epoch
(or redshift $z_f$) of halo formation, what allows us to
quantify correlation of halo parameters with the process
of growth of perturbations and halo formation.

Evidently the basic assumptions of the simple model
(\ref{rmod}) are not realistic. Indeed the processes
of merging and anisotropic matter accretion unpredictably
changes the density. Analysis of high resolution
simulations (Demia\'nski et al. 2011) shows that
influence of these random factors together with the
strong anisotropy of the early period of DM halo
formation depends upon the halo mass and, for example,
it is moderate for clusters of galaxies. It is accepted
that for the simple $\Lambda$CDM cosmological model and
for more massive objects the expression (\ref{rmod})
describes reasonably well both the observations and
simulations. However, the formation of low mass objects
is regulated by other factors and thus their structure
cannot be described by the same relations. These comments
attempt to explain the main reasons why the model
(\ref{rmod},\,\ref{mvir}) has limited applicability.

In order to describe the complex process of DM halo formation
in a wide range of virial masses of objects we use the more
general {\it phenomenological} relation
\be
\langle\rho_{vir}\rangle=18\pi^2\Phi(M_{vir})\langle
\rho_m(z_f)\rangle\approx\rho_{200}\Phi(M_{vir})(1+z_f)^3\,,
\label{pmod}
\ee
where $\Phi(M_{vir})\geq 1$ is a smooth slowly varying
function of $M_{vir}$. This relation preserves the
universality of (\ref{rmod}) and provides an unified
self consistent description of observed properties of DM
halos in a wide range of masses $10^6\leq M_{vir}/M_\odot\leq
10^{15}$\,. In particular, it reproduces the observed weak
mass dependence of the redshift $z_f$ and the density
$\langle\rho_{vir}\rangle$ for both clusters of galaxies
and low mass THINGs, LSB, UDG and dSph galaxies. Further
discussion of these problems can be found below in sections
3, and 4\,.

Precise  observations of DM periphery of both clusters and
galaxies are problematic  owing to the strongly irregular
matter distribution in their outer regions.  Hence, we use only
estimates of the more stable parameter - the virial mass of object
as its leading characteristic. For example, sometimes the relation
(\ref{rmod}) is used for description of clusters of galaxies
under the arbitrary assumption that the cluster is formed at
the observed redshift, $1+z_f\equiv 1+z_{obs}$. In this case
the relation (\ref{rmod}) allows to determine formally the mean
virial density $\langle\rho_{vir}\rangle$ and, for a given mass
of cluster $M_{vir}$, its virial radius, $R_{vir}$. However, this
result is evidently incorrect because, in fact, we can only
conclude that $z_f\geq z_{obs}$, what is trivial for galaxies.
For clusters the difference between $1+z_f$ and $1+z_{obs}$ can
be as large as $\sim 1.5 - 2$.

More stable and refined method to determine the redshift
of halo formation uses characteristics of the halo core
rather than its periphery. In this case  we use the
following expression for the concentration
\be
c(M_{vir},z_f)\approx 0.12M_{12}^{1/6}(1+z_f)^{7/3},\quad
M_{12}=\frac{M_{vir}}{10^{12}M_\odot},
\label{cmz}
\ee
(Demia\'nski\,\&\,Doroshkevich 2014). Together
with relations (\ref{nfw-ms}) and (\ref{pmod}) we get for the mean
density of the central core
\be
\langle\rho_s\rangle=\rho_{cc}M_{12}^{1/2}(1+z_f)^{10}\Phi(M_{vir})\,,
\label{cdns}
\ee
\[
\rho_{cc}=0.38\cdot 10^{-3}\frac{\rho_{200}}{f_m(c)}
\approx\frac{2.5}{f_m(c)}\frac{M_\odot}{kpc^3}=
\frac{1.9}{f_m(c)}\cdot 10^{-31}\frac{g}{cm^3}\,.
\]
This relation links the redshift $z_f$ with the virial mass
$M_{vir}$ and the mean density of halo core $\langle \rho_s
\rangle$ and allows to determine $z_f$. Application of this
approach requires additional observations. However, it is less
sensitive to random deviations of characteristics of periphery
of objects. Moreover for DM dominated high density objects
observed at $z_{obs}\ll 1$ (such as the dSph galaxies)
determination of the redshift $z_f$ through the parameters of
the central core is also preferred.

The main weakness of this approach is the possible impact of
baryonic component that can be specially important for
galaxies with moderate DM domination. However, for dSph galaxies
this effect can be comparable with the uncertainties in
measurements of parameters, what is clearly seen, when one
compares the results presented in Walker et al. (2009) and
Kirby et al. (2014) (see section 4).

Comparison of (\ref{pmod}) and (\ref{cdns}) indicates
that the density of halo core $\langle\rho_s\rangle$
is more sensitive than $ \langle\rho_{vir}\rangle $ to both
the virial mass and the redshift of
formation. However, for all observed samples (section 4) $z_f$
is correlated with $M_{vir}$:
 \be
\eta_f=(1+z_f)M_{12}^{0.077}\approx 4.1(1\pm 0.1)\,,
 \label{zm}
\ee
(see also Demia\'nski\,\&\,Doroshkevich 2014). Allowing
for this correlation we see from (\ref{cdns}) that actually
$\langle\rho_s\rangle$ is a weak function of both the virial
mass $M_{vir}$ and the redshift of formation $z_f$.

\subsection{Excursion set approach and shape of the power
spectrum}
It is well known that at redshifts
$z\geq 3$ the formation of galactic scale halos dominates,
but the typical mass of halos increases with time and massive
clusters of galaxies are formed later at redshifts
$z\leq 2$. This correlation between the redshift of halo
formation $z_f$ and the halo mass $M_{vir}$ is described by
current models of halo formation (Press \& Schechter 1974;
Bardeen et al. 1986; Bond et al. 1991; Sheth \& Tormen 2002;
2004). They are based on the excursion set approach applied
to the initially Gaussian random density perturbations. They
reduce description of characteristics of the formed halos to
the problem of crossing of an appropriate barrier by particles
undergoing Brownian motion. A close link between the power
spectrum and properties of DM halos and Large Scale Structure
had been demonstrated in many simulations.

All theoretical models of halos formation predict that the
distribution functions of halo characteristics are dominated
by a typical Gaussian term:
\be
dP(M_{vir})\propto \exp[-\alpha\Psi^2(M_{vir})]dM_{vir}\,,
\label{sz}
\ee
\[
\Psi(M_{vir})=\sigma_m(M_{vir})B(z_f(M_{vir}))\,,
\]
where $B(z_f)$ (\ref{bzz}) describes
the growth of density perturbations and $\sigma_m$
is the dispersion of density perturbations given by
\be
\sigma_m^2(M)=\frac{1}{2\pi}\int_0^\infty k^2p(k)W^2(k,M) dk\,,
\label{sigm}
\ee
\[
W(x)=3(sin\,x/x^3-cos\,x/x^2),\quad x=kr\propto kM^{1/3}\,.
\]
Here $p(k)$ is the power spectrum and $W$
is the Fourier transform of the real-space top-hat filter
corresponding to a spherical mass $M$. Thus, the function
$\Psi(M_{vir})$ characterizes the amplitude of perturbations
with mass $M_{vir}$. The usually used condition
\be
\Psi(M_{vir})=\sigma_m(M_{vir})B(z_f(M_{vir}))\approx const\,,
\label{const}
\ee
provides the expected approximate self similarity of
the process of halo formation and progressive growth with
time of the virial mass of halos.

For the CDM -- like power spectrum and for the standard
normalization of perturbations on $\sigma_8$ the function
$\sigma_m$ and $\alpha$ are well fitted by the following
expressions
\be
\sigma_m=\frac{3.31\sigma_8 M_{12}^{-0.077}}{1+0.177M_{12}^{0.133}+
0.16M_{12}^{0.333}},\quad \alpha=\frac{1.686^2}{2\sqrt{2}}
\approx 1\,,
\label{sig_cdm}
\ee
where $M_{vir}= M_{12}10^{12}M_\odot$, 1.686 is the critical
overdensity (height of the barrier) and $\sigma_m\approx
\sigma_8$ for
\[
M_{12}=\frac{4\pi}{3}\frac{\langle\rho_m\rangle}{10^{12}M_\odot}
\left(\frac{8Mpc}{h}\right)^3\approx 210
\Theta_m\left(\frac{0.7}{h}\right)^3\,.
\]
It is interesting that for $M_{12}\leq 1$ we get from
(\ref{bzz}), (\ref{zm}), and (\ref{sig_cdm}) that
\be
\Psi(M_{vir})\approx \frac{4.47\sigma_8}{(1+z_f)M_{12}^{0.077}}=
1.1\sigma_8\left(\frac{4.1}{\eta_f}\right)\approx const\,,
\label{par3}
\ee
what is consistent with (\ref{const}) and
demonstrates that properties of low mass objects are
in agreement with the CDM--like shape of the small scale
power spectrum. We will discuss in more details the
observed properties of the function $\Psi(M_{vir})$ in
section 5.

\section{Expected properties of the relaxed DM halos}

We consider the DM halos as a one parametric sequence of
objects all properties of which depend upon their virial mass.
This means that we consider all DM halos as similar ones.
Together with the virial characteristics of halos, namely,
the mass, $M_{vir}$, radius $R_{vir}$, and density
$\langle\rho_{vir}\rangle$ we also consider the mean
characteristics of their central cores, namely, the
density $\langle\rho_s\rangle$, pressure, $\langle
P_s\rangle$, temperature $\langle T_s\rangle$ and entropy
$\langle S_s\rangle$. The very important characteristic
of a halo is the redshift of its formation, $z_f$, that
was introduced by (\ref{pmod}), it approximately
characterizes the end of the period of halo formation
and relaxation.

The DM temperature and velocity dispersion $\sigma_v$ in the
core usually are not observed, but within relaxed DM cores
$\sigma_v$ is close to the circular velocity, $v_c(r)$,
(Demia\'nski\,\&\,Doroshkevich 2014) \be \sigma_v^2(r)\approx
v_c^2(r)\sqrt{r_s/r}\,. \label{tc} \ee Because of this we can
estimate the expected central parameters of a DM halo as:
\[
\langle T_s\rangle\approx \frac{m_{b}v_c^2}{2}\approx
1.8eVM_{12}^{5/6}(1+z_f)^{10/3}(\Phi/f_m)^{1/3}\mu_{DM}\,,
\]
\be
\langle P_s\rangle\approx 10^{-7}eV/cm^3M_{12}^{4/3}
(1+z_f)^{40/3}(\Phi/f_m)^{4/3}\,,
\label{ps}
\ee
\[
\langle S_s\rangle\approx 80 cm^2keV \frac{M_{12}^{1/2}}
{(1+z_f)^{10/3}}\mu_{DM}^{5/3}(f_m/\Phi)^{1/3}\,.
\]

Using the correlation between $z_f$ and $M_{vir}$ (\ref{zm})
we finally get
\[
\langle\rho_s\rangle\approx \eta_\rho M_{12}^{-0.2}
\frac{\Phi}{f_m}\left(\frac{\eta_f}{4.1}\right)^{10}\,,
\]
\be
\langle P_s\rangle\approx \eta_p M_{12}^{0.4}\left(
\frac{\Phi}{f_m}\right)^{4/3}\left(\frac{\eta_f}{4.1}
\right)^{40/3}\,,
\label{psm}
\ee
\[
\langle S_s\rangle\approx \eta_s M_{12}^{0.73}\left(\frac{\eta_f}{
4.1}\right)^{-10/3}\left(\frac{f_m}{\Phi}
\right)^{1/3}\mu_{DM}\,,
\]
  where expected values of the constants are
\be
\eta_\rho\approx 2.6\cdot 10^6 \frac{M_\odot}{kpc^3},\quad
\eta_p\approx 22\frac{eV}{cm^3},\quad \eta_s\approx 0.9cm^2keV\,.
\label{e-etas}
\ee

The expected self similar character of the internal
structure of DM cores is manifested by a weak dependence
of the parameters $\eta_f,\,\eta_\rho,\,\eta_p,\,\&\,\eta_s$
upon the virial mass of objects in a wide range of masses.
For the observed objects values of these parameters are
obtained in the next section 4. For the best sample of 19
dSph galaxies with $M_{vir}\leq 10^9M_\odot$ and 19
CLASH clusters with $M_{vir}\geq 10^{14}M_\odot$ we
have
\be
\eta_\rho\approx 1.3\cdot 10^6(1\pm 0.9)\frac{M_\odot}{kpc^3},
\quad\eta_s\approx 1.2(1\pm 0.5)cm^2keV,\quad
\eta_f\approx 4.1(1\pm 0.1)\,,
\label{s38}
\ee
what is quite similar to expectations (\ref{e-etas}).  Here
and below we use the correction factor
\be
\Phi(M_{vir})=(1+M_f/M_{vir})^{0.3},\quad M_f\approx 8\cdot
10^{12}M_\odot\,,
\label{phi_m}
\ee
that allows us to obtain an unified self consistent description
of both discussed galaxies and
clusters of galaxies. Thus, for larger masses $M_{vir}\gg M_f$,
$\Phi\rightarrow 1$, and expressions (\ref{rmod}) and
(\ref{pmod}) become identical to each other. On the other hand,
in the opposite case $M_{vir}\ll M_f$ the function
(\ref{phi_m}) allows to reconcile  theoretical expectations
(\ref{psm}) with observations (\ref{ffit}) presented in
section 4. For low mass dSph galaxies this approach decreases
the redshift of formation $z_f$ down to values consistent
with the observed age of stars and Planck estimates of the
redshift of reionization.

Of course the function $\Phi(M_{vir})$ (\ref{phi_m}) should be
considered only as the first approximation. But to
obtain more refined and justified description of the redshift
of formation $z_f$ (\ref{pmod}) and DM halo parameters
(\ref{psm}) we need to have both richer observational data with
only moderate scatter and corresponding high resolution
simulations.

As is seen from (\ref{zm}) and (\ref{cmz}) the so defined
concentration is only weakly depended on the virial mass,
\be
\langle c\rangle\approx 3(\eta_f/4.1)^{7/3}M_{12}^{0.004}\,.
\label{cc}
\ee

\section{Observed characteristics of DM dominated galaxies and
clusters of galaxies}

For our analysis we used more or less reliable observational
data for $\sim 19$ DM dominated clusters of galaxies, one
UDG and $\sim 30$ dSph galaxies. For comparison we use also
observations of 30 groups of galaxies and $\sim 11$ THINGS
and LSB galaxies. In this section we consider properties of
the central cores of virialized DM halos using the
approximations summarized in the previous sections. We
characterize halos by their virial mass $M_{vir}$ and redshift
of formation, $z_{f}$. We assume that at $z\leq z_{f}$ the halos
mass and core temperature and density do not change
significantly.

\subsection{The CLASH clusters}

For 19 clusters of the CLASH sample (Merten et al. 2015) in
addition to the usually presented observed virial mass of
clusters, $M_{vir}$, there are also estimates of the size
and density of central core, $r_s,\,\&\,\rho_0$. For this
survey the published parameters $R_{vir},\,\&\,\rho_{vir}$ are
obtained under the assumption that $z_f=z_{obs}$ and they are
not used in our analysis. The virial mass of
clusters, $M_{vir}$, presented in Merten et al. (2015) is
close to the estimates of $M_{vir}$ obtained by Umetsu et
al. (2014), what confirms reliability of these estimates.

For these clusters parameters of cores can be used without
serious corrections. Main results are plotted in Figs.
\ref{fprs} and \ref{sig6}. We get for this survey
\be
\langle c\rangle\approx 3.7(1\pm 0.2),\quad
\langle 1+z_f\rangle\approx 2.3(1\pm 0.1),
\quad \langle\eta_f\rangle\approx 3.9(1\pm 0.1)\,,
\label{thet-20}
\ee
\[
\langle\rho_s\rangle\approx 5\cdot 10^5(1\pm 0.5)
M_\odot/kpc^3,\quad \langle\eta_\rho\rangle\approx 1.6\cdot
10^6(1\pm 0.5)M_\odot/kpc^3,
\]
\[
\langle S_s\rangle\approx 160(1\pm 0.9)cm^2keV\,,
\quad \langle\eta_s\rangle\approx 1.2(1\pm 0.1)cm^2keV\,,
\]
\be
\langle M_{12}\rangle\approx 10^3,\quad
\langle \Psi(M)\rangle\approx 0.4\sigma_8(1\pm 0.1)\,.
\label{par1}
\ee

\subsection{Ultra diffuse galaxies}

Now there are new possibilities of observations of the Ultra
Diffuse Galaxies (UDG) (
Liu et al. 2015; Beasley et al. 2016: Martinez-Delgado et
al. 2016). For one of this object -- DM dominated galaxy
Dragonfly 44 (van Dokkum et al. 2016) -- there are
observations of $M_{1/2},\,\,r_{1/2},\,\&\,\sigma_v$ what
allows us to repeat the analysis performed for the dSph
galaxies. Thus, we get for this galaxy
\be
c=2.8\quad 1+z_f\approx 5.2,\quad \eta_f\approx 3.8,\quad
M_{vir}\approx 1.4\cdot 10^{10}M_\odot,\,.
\label{udg}
\ee
\[
\rho_s\approx 4.2\cdot 10^{8}M_\odot/kpc^3,\quad
\eta_\rho\approx 2.4\cdot 10^7M_\odot/kpc^3\,,
\]
\[
S_s\approx 0.078cm^2 keV,\quad \eta_s\approx 3.3cm^2keV,
\quad \Psi(M)\approx 1.1\sigma_8\,,
\]

These data are presented in Figs. \ref{fprs}\&\,\ref{sig6}.
As is seen from these Figures this galaxy is located
halfway between the CLASH clusters and dSph galaxies and
quite well fits with other data plotted in these Figures.
This can be considered as the independent evidence in
favor of our approach and inferences. We hope that further
investigations of UDG galaxies will improve our results.

\subsection{Two catalogues of the dSph galaxies}

Recently the main observed parameters the dSph
galaxies were listed and discussed in many papers. Published
characteristics of these galaxies vary from paper to paper
and are presented with significant scatter. From samples
presented in Walker et al. (2009 \& 2011); Collins et al.
(2014); Tollerud et al. (2012 \& 2014) we selected 19
objects with high DM density, what suggests that these
objects were formed at high redshifts and can be considered
as samples of earlier galaxies responsible for reionization.
This means that for these galaxies the redshift of formation
$z_f$ should be comparable with redshift of reionization
$z_{reio}\sim 9$ determined by WMAP and Planck missions.
For these galaxies we find
\be
\langle c\rangle\approx 3.7(1\pm 0.2),\quad \langle
1+z_f\rangle\approx 9.9(1\pm 0.1),\quad
\langle\eta_f\rangle\approx 4.2(1\pm 0.1)\,,
\label{thet-wal}
\ee
\[
\langle\eta_\rho\rangle\approx 2.2\cdot 10^5(1\pm 0.7)M_\odot/kpc^3,
\quad \langle\eta_s\rangle\approx 0.9(1\pm 0.3)cm^2keV\,,
\]
\[
10^6\leq M_{vir}/M_\odot\leq 10^8,\quad
\langle\Psi(M)\rangle \approx 1.1(1\pm 0.1)\sigma_8\,.
\]
 Other results for this survey are plotted in Figs 1 \& 2.

Second catalog of dSph galaxies (Kirby et al. 2014) contains
30 satellites of Milky Way and Andromeda with independent
measurements of their parameters. Some of them
significantly differ from those presented in Walker et al.
(2009 \& 2011) and Tollerud et al. (2012 \& 2014). We have
for this survey
\be
\langle c\rangle\approx 3.3(1\pm 0.2),\quad \langle
1+z_f\rangle\approx 8.1(1\pm 0.1),\quad
\langle\eta_f\rangle\approx 4.1(1\pm 0.1)\,,
\label{thet-krb}
\ee
\[
\langle\eta_\rho\rangle\approx 1.5\cdot 10^5(1\pm 0.7)
M_\odot/kpc^3,\quad \langle\eta_s\rangle\approx
1.1(1\pm 0.4)cm^2keV\,,
\]
\[
10^6\leq M_{vir}/M_\odot\leq 10^8,\quad
\langle\Psi(M)\rangle\approx 1.1(1\pm 0.1)\sigma_8\,.
\]
Other results for this survey are plotted in Figs 1 \& 2.
This comparison of results (\ref{thet-wal}) and (\ref{thet-krb})
allows to verify representativity and reliability of obtained
estimates.

\subsection{The groups of galaxies}

For comparison parameters of 30 suitable groups of galaxies
with $z_{obs}\ll 1$ are taken from the catalog of
Makarov \& Karachentsev (2011). For these objects we cannot
identify  the central core. But these data allow us to
estimate -- with large scatter -- the mean density and
the redshift of formation as
\[
1.2\leq M_{12}\leq 200,\quad 100kpc\leq R_{vir}\leq 700kpc,
\quad \langle\rho_{vir}\rangle\approx 2.7\cdot 10^5(1\pm
0.9)M_\odot/kpc^3\,,
\]
\be
\langle 1+z_f\rangle\approx 3.1(1\pm 0.3),\quad
\langle\eta_f\rangle\approx 3.1(1\pm 0.3),\quad
\langle \Psi(M)\rangle\approx 0.74\sigma_8(1\pm 0.3)\,,
\label{grup1}
\ee
and to fill partly the empty region in Fig. \ref{sig6}
between the galaxies and clusters of galaxies. Main results
of our analysis are plotted in Fig. \ref{sig6}.

\subsection{The THINGS and LSB galaxies}

For 11 LSB and THING galaxies (de Blok et al. 2008; Kuzio
de Naray et al. 2008) the observed
rotation curves are measured up to large distances, what
allows us to estimate, with a reasonable reliability,
the virial masses $M_{vir}$ and the mean virial density,
$\langle\rho_{vir}\rangle$, and, finally, using the relation
(\ref{pmod}) to estimate the redshift of formation and the
virial mass of these objects.  we get
\be
\langle 1+z_f\rangle\approx 2.3(1\pm 0.3),\quad
\langle\eta_f\rangle\approx 2.8(1\pm 0.3),\quad
\langle M_{12}\rangle\approx 0.24(1\pm 0.8)\,,
\label{galac}
\ee
\[
\langle\rho_{vir}\rangle\approx 6.2\cdot 10^5(1\pm 0.9)M_\odot/kpc^3,
\quad\langle\eta_\rho\rangle\approx 0.6\cdot 10^5(1\pm 0.9)
M_\odot/kpc^3\,,
\]
\[
\langle \Psi(M)\rangle\approx 1.6\sigma_8(1\pm 0.2)\,.
\]

\begin{figure}
\centering
\epsfxsize=7. cm
\epsfbox{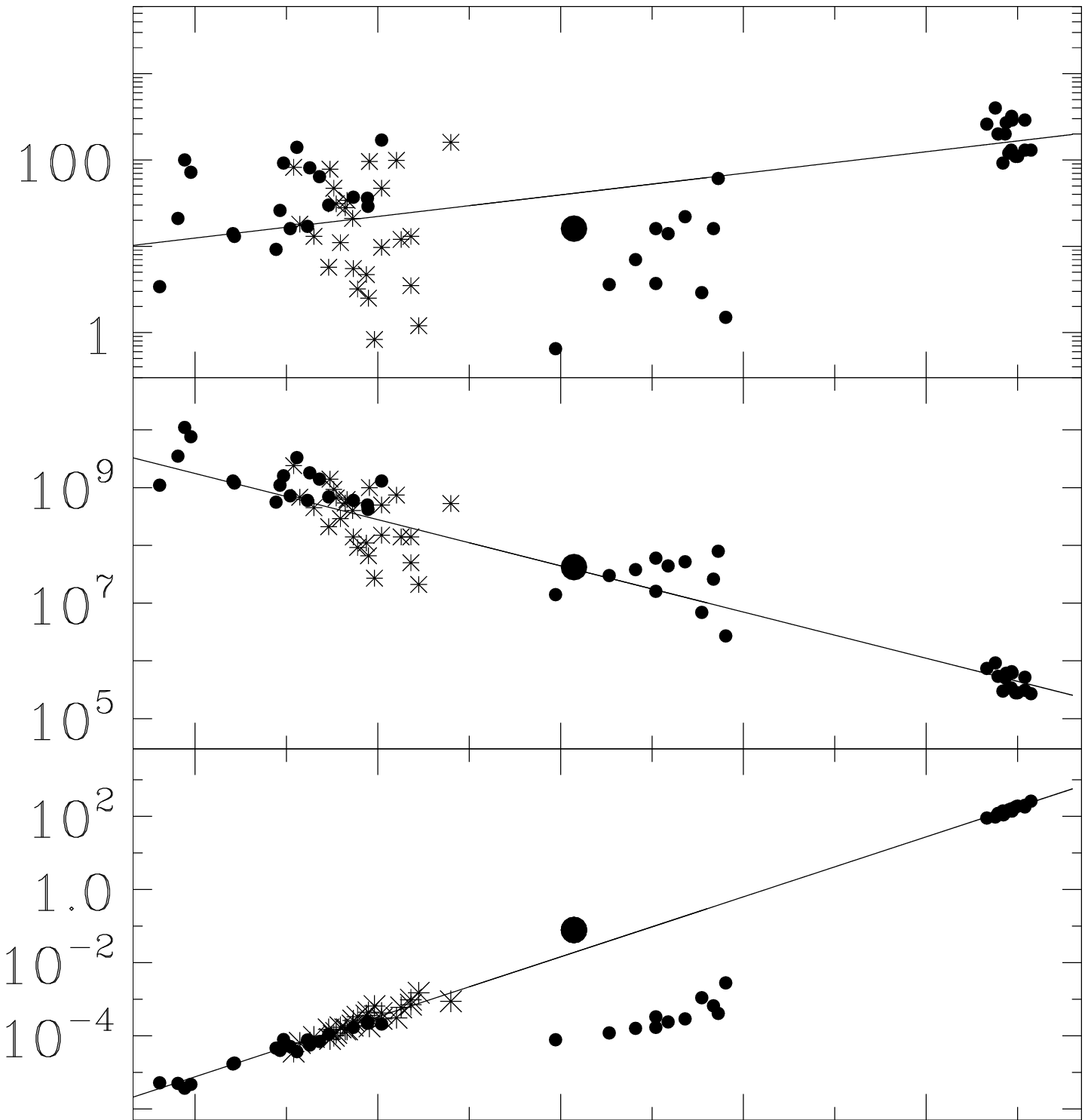}
\vspace{0.7cm}
\caption{The functions $P_s(M_{vir}),\,\rho_s(M_{vir}),\,
\&\,S_s(M_{vir})$ are plotted for the two samples of dSph
galaxies -- (\ref{thet-wal}) left points and (\ref{thet-krb})
stars, THING and LSB galaxies (central points) and CLASH
clusters of galaxies (right points). For the Dragonfly 44
values (\ref{udg}) are plotted by filled dark circle. Fits
(\ref{ffit}) are plotted by solid lines.
}
\label{fprs}
\end{figure}

The complex internal structure of these galaxies
and the significant influence of stars, discs and diffuse
baryonic component restricts the number of objects for
which reasonable DM characteristics can be derived. For
11 galaxies results of our analysis are presented in
 Figs. \ref{fprs} and \ref{sig6}.

\subsection{Basic characteristics of the DM cores}

For two samples of dSph galaxies, Dragonfly 44 galaxy, and
CLASH clusters of galaxies parameters of their DM cores
(the pressure, $P_s$, density, $\rho_s$, and entropy,
$S_s$) are plotted in Fig. \ref{fprs} and are fitted
by the expressions:
\[
\langle P_s\rangle=140(1\pm 0.8)eV/cm^3M_{12}^{0.1}\,,
\]
\be
\langle \rho_s\rangle=10^7(1\pm 0.6)\frac{M_\odot}{kpc^3}
M_{12}^{-0.45}\,,
\label{ffit}
\ee
\[
\langle S_s\rangle=0.66(1\pm 0.4)keV cm^2 M_{12}^{0.85}\,,
\]
\[
\langle 1+z_f\rangle=4.1(1\pm 0.17) M_{12}^{-0.077}\,,
\]
where again $M_{12}=M_{vir}/10^{12} M_\odot$. For comparison
in Fig. \ref{fprs} the same functions are also plotted
for THING and LSB galaxies. Large deviations of these
parameters from the fits are caused by uncertainties in
estimates of the influence of baryonic component and
other parameters of the DM cores. Large
scatter of the functions (\ref{ffit}) reflects mainly the
large scatter of the observational data and the natural
random variations of object characteristics.

\section{Shape of the power spectrum}

Observations of the relic microwave background radiation
allow to determine the shape of the initial power spectrum
of density perturbations at large scales starting from
clusters of galaxies (Komatsu et al. 2011; Ade et al. 2016).
However, information about the power spectrum at small scale
and composition and properties of dark matter is still
missing. In this section we consider in more details the
function $\Psi(M_{vir})$ introduced by (\ref{sz}) and link
its mass variations with the shape of the power spectrum.

\subsection{Theory versus observations}

Both simulations and observations show that characteristics
of the DM dominated halos are much more stable than
characteristics of the baryonic component, and they can be
used to characterize the small scale power spectrum of
density perturbations. Indeed our analysis performed in
sections 3\,\&\,4 shows that we can find a one  to  one
correspondence between the observed parameters of the DM
halos and the redshift of object formation,
$z_{f}$. According to the present day models of halo
formation (Press, Schechter, 1974; Peebles 1980; Bardeen
et al. 1986; Bond et al. 1991; Sheth\,\&\, Tormen 2002,
2004) the variations of redshift $z_f$ with the virial
mass of created objects characterize both the power
spectrum of density perturbations and the real period
of objects formation. As was discussed above in these
models the redshift of object formation, $z_f$, is a weak
function of virial mass. Therefore we can expect that the
function $\Psi(M_{vir})$ describing this process shows
also weak dependence upon the virial mass. However,
estimates (\ref{par1}) -- (\ref{galac}) demonstrate
unexpectedly significant mass dependence  of this function.
\[
 \langle\Psi(M)\rangle/\sigma_{8}\approx 0.4,\,0.74,\,1.6,
\,1.1,\,1.1,\,1.1 \,.
\]
This effect suggests a possible deviation of the small scale
power spectrum from that observed at larger scales by the
WMAP and Planck missions and now accepted in the $\Lambda$CDM
cosmological model.

\begin{figure}
\centering
\epsfxsize=7.cm
\epsfbox{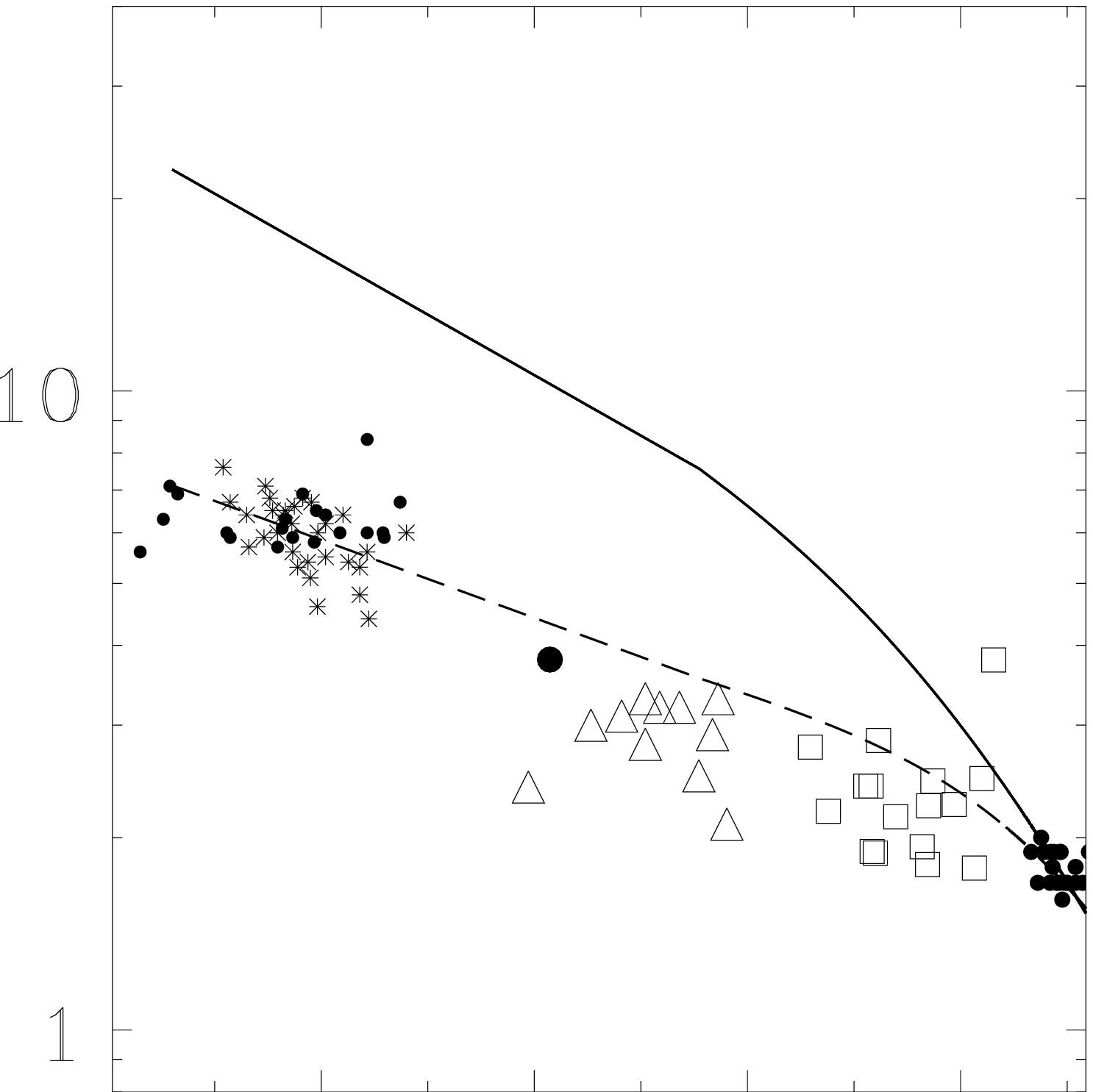}
\vspace{0.7cm}
\caption{Dispersion of the density perturbations $\sigma_m^*
=\sigma_m(M)/0.43$ (\ref{sigm},\,\ref{s**}) for the standard
$\Lambda$CDM power spectrum (\ref{sig_cdm}) and the combined
spectrum (\ref{best1}) are plotted by solid and long dashed
lines vs. $M_{vir}/M_\odot$. Function $B^{-1}(M_{vir})$ (\ref{bbz})
is plotted for the two samples of dSph galaxies (left group of
points and stars), for Dragonfly 44 galaxy (filled circle) and
for CLASH clusters of galaxies (right group of points). For
THINGS and LSB galaxies and for groups of galaxies the
functions $B^{-1}(M_{vir})$ are plotted by triangles and squares.
}
\label{sig6}
\end{figure}

This effect is illustrated in Fig. \ref{sig6} where the
function $B^{-1}(M_{vir})$ (\ref{bbz}) is plotted for the
samples of CLASH clusters, groups of galaxies, THING and
LSB galaxies and for two samples of dSph galaxies.
These functions are compared with the dispersion of
the density perturbations $\sigma_m(M_{vir})$ (\ref{sigm})
calculated for the standard $\Lambda$CDM power spectrum
(\ref{sig_cdm}). To more clearly represent the trend we plot
in Fig. \ref{sig6} the function
\be
\sigma_m^*=\sigma_m(M_{vir})/0.43\,, \label{s**}
\ee
that satisfies  the condition
\be
\Psi^*(M_{vir})=\sigma_m^*(M_{vir})B(z_f(M_{vir}))\approx 1\,,
\label{correc}
\ee
for the CLASH clusters of galaxies.
The strong differences between observations and expectations
of the standard CDM--like power spectrum are clearly seen
in Fig. \ref{sig6}.

At the same figure the observed points $B^{-1}(M_{vir})$ are
well fitted by functions $\sigma_m^*=\sigma_m/0.43$
obtained for the more complex power spectrum
\be
p_m(k)=0.1p_{cdm}(k)+0.9 p_{wdm}(k),\quad
\sigma_m^2=0.1\sigma_{cdm}^2+0.9\sigma_{wdm}^2\,.
\label{best1}
\ee
Here $p_{cdm}(k)$ is the standard CDM power spectrum. The
function
\be
\sigma_{wdm}\approx 1.3\sigma_8/(1+0.05M_{12}^{0.4})\,,
\label{pwdm}
\ee
corresponds to the power spectrum describing the contribution
of the large scale perturbations only. The three functions
$\sigma_m^*(M_{vir})$ plotted in Fig. \ref{sig6} are identical
to each other for $M_{vir}\geq 10^{15}M_\odot$.

As an example of the power spectrum with damped small scale
part (\ref{pwdm}) we use the power spectrum of WDM particles
(Viel et al. 2013), namely,
\be
p_{wdm}(q)\approx p_{cdm}(q)[1+(\alpha_w q)^{2.25}]^{-4.46}\,,
\label{viel}
\ee
\[
q=\frac{k}{\Omega_mh^2},\quad
\alpha_w=6\cdot 10^{-3}\left(\frac{\Omega_mh^2}{0.12}
\right)^{1.4}\left(\frac{1 keV}{m_w}\right)^{1.1}\,,
\]
where $m_w\sim (50 - 100) eV$ and the comoving wave number
$k$ is measured in $Mpc^{-1}$. This mass of  $m_w$ corresponds
to the damping scales $M_{dmp}$ and $D_{dmp}$
\be
\sigma_{wdm}(M_{dmp})=0.5\sigma_{wdm}(0),\quad
M_{dmp}=1.8\cdot 10^{15}M_\odot,\quad D_{dmp}\sim (10 - 20)
h^{-1}Mpc  \,.
\label{md}
\ee

Let us note however that in $p_{wdm}$ and $\sigma_{wdm}$ the
mass $m_w$ is only a  formal parameter allowing to introduce
the suitable damping scales $M_{dmp}$ and $D_{dmp}$ in the
function $\sigma_{wdm}$ (\ref{pwdm}) and such particles
need not really exist. The function $\sigma_{wdm}$
(\ref{pwdm}) is weakly sensitive to the shape of the
spectrum (\ref{viel}), when the suppression of power
occurs sufficiently rapidly. For example, the spectrum
with the Gaussian damping
\be
p_{wdm}(q)\approx p_{cdm}(q)[1+\exp(q^2/q_{dmp}^2)]^{-2}\,,
\label{gauss}
\ee
and a suitable value $q_{dmp}$ provides the same $\sigma_{wdm}$
(\ref{pwdm}). At the same time the CDM--like shape of the
small scale power spectrum is preserved in (\ref{best1}),
 what is consistent with similarity of the
expressions (\ref{sig_cdm}) and (\ref{zm}) for $M_{12}\leq 1$.
It is apparent that such decrease of the amplitude of small
scale perturbations eliminates discrepancy between estimates
of the function $\Psi(M_{vir})$.

\subsection{'Too Big To Fail' approach}

Essential support for our inferences comes from comparison
of the circular velocities of simulated low mass galaxies
and observed dSph satellites of Milky Way and Andromeda
(Boylan--Kolchin et al. 2012; Garrison--Kimmel et al.,
2014a,b; Tollerud et al. 2014; Hellwing et al. 2015; Brook,
\& Cintio 2015). It demonstrates that the
circular velocities of objects in simulations performed
with the standard $\Lambda$CDM power spectrum reproduce
observations for more massive objects, but regularly
overestimate the observed circular velocities for less
massive objects.

It seems that this discrepancies can be related to
efficiency of the environmental processes and complex
evolutionary history in the context of the hierarchical
formation of objects. This  problem is discussed by Wetzel
et al. (2015). But the similar effects are observed also
for galaxies in the Local Group, where, in particular,
there is the deficit of low mass galaxies in observations
as compared with predictions of the standard simulations
(Klypin et al. 2015). However, simulations with the standard
WDM power spectrum do not reproduce observations.

These results show that more complex improvements
of the standard theoretical models are required. But
these indications of qualitative disagreements between
standard simulations and observations are not yet followed
by quantitative estimates of required corrections.

\section{Discussions and conclusions}

In this paper we  propose a new approach that could solve
some important problems of modern cosmology:
\begin{enumerate}
\item{} Explain the self similarity of the internal structure
of virialized DM dominated objects in a wide range of masses,
what is manifested as the regular dependence of the
central pressure, density, entropy and the epoch of DM halos
formation (\ref{psm}, \ref{ffit}) on the virial mass of
objects.
\item{} Explain unexpectedly weak dependence of the DM pressure
 in halos on their virial masses.
\item{}Establish the real composition of dark matter
and the shape of the small scale power spectrum of density
perturbations\,.
\end{enumerate}

Summing up let us note that the proposed approach allows us to
consider and to compare properties of the observed DM
dominated objects in an unprecedentedly wide range of masses
$10^6\leq M_{vir}/M_\odot\leq 10^{15}$. The main results
 plotted in Fig. \ref{fprs}
unexpectedly favor the regular self similar character of the
internal structure of these objects, what confirms the main
expectations of the NFW model and the regular character of
the power spectrum without sharp peaks and deep troughs.

On the other hand as is seen from Fig. \ref{sig6} our
results favor the models with more complex power spectrum
with significant excess of power at cluster scale and/or,
alternatively, deficit of power at dwarf galactic scales.
Both alternatives seem to be quite important and call for
more detailed observational study of DM dominated objects.
Our results supplement the traditional investigations of
galaxies at high redshifts (see, e.g., Bouwens et al., 2015;
Ellis et al. 2016).

Of course these results cannot be considered as a strong
indication that the standard $\Lambda$CDM model should be
modified, but they demonstrate again some intrinsic problems of
this model. The strong links of DM characteristics of observed
relaxed objects with the initial power spectrum and cosmological
model is well known. Results presented in Figs. \ref{fprs} and
\ref{sig6} demonstrate the regular correlations of properties
of DM component with the virial masses of the relaxed objects.
These correlations can be successfully interpreted in the
framework of the standard model of halo formation based on the
excursion set approach, but applied to a more complex power
spectrum. However, these unexpected results should be tested with
suitable set of high resolution simulations before their real
status could be accepted.

Of course the observational base used in our discussion is
very limited and it should be extended by more observations
of objects with masses $M\leq 10^{12} M_\odot$, what may be
crucial for determination of the shape of the initial power
spectrum, the real composition of dark matter and even the
models of inflation. Unfortunately more or less reliable
estimates of the redshift $z_{f}$ can be obtained mainly for
the DM dominated objects or for objects with clearly
discriminated impact of DM and baryonic components.
The most promising results can be obtained for
the mentioned in section 4 population of Ultra Diffuse
Galaxies now represented by the galaxy Dragonfly 44 only
(see Figs 1\,\&\, 2). We hope that the list of possible
appropriate candidates will be extended.

Here we consider a phenomenological model of the power
spectrum that is composed of fraction $g_{cdm}\sim 0.1 -
0.2$ of the standard CDM spectrum and fraction $g_{wdm}\sim
0.9 - 0.8$ of the HDM spectrum with the transfer function
(\ref{viel}). Unexpectedly in the spectrum (\ref{best1}) the
contribution of low mass particles with relatively large
damping scale (\ref{md}) dominates, what can be considered
as reincarnation in a new version of the earlier rejected
HDM model. Further progress can be achieved with more
complex models of the power spectrum and/or more realistic
transfer functions instead of (\ref{viel},\, \ref{gauss}),
what implies also more realistic complex composition of the
DM component. However, all such problematic multi parametric
proposals should be considered in the context of the general
cosmological and inflationary models.

\subsection{Acknowledgments}
This paper was supported in part by  the grant of the
President of RF for support of scientific schools
NSh-6595.2016.2. We thank S. Pilipenko, A. Klypin for
useful comments.

\end{document}